\documentclass[pre,twocolumn,twoside,showkeys,floatfix]{revtex4}
\usepackage{dcolumn}%
\usepackage{bm}%
\usepackage{graphicx,textcomp}
\usepackage{amssymb,amsfonts,amsmath}
\usepackage{color}
\usepackage{nicefrac}
\usepackage{algorithm}  
\usepackage{algorithmicx}  
\usepackage{algpseudocode}  

\begin{document}

\title{How initial condition impacts aggregation -- a systematic numerical study}

\author{Micha\l{} \L{}epek\footnote{e-mail: lepek@if.pw.edu.pl}}

\affiliation{Warsaw University of Technology, The Faculty of Physics, Koszykowa 75, Warsaw, Poland, PL-00-662}

\date{\today}


\begin{abstract}
In a process of aggregation, a finite number of particles merge irreversibly to create growing clusters. In this work, impact of particular initial conditions: monodisperse, power--law, exponential, and inspired by condensation nuclei was tested against several aggregation processes: constant, additive, product, electrorheological, anti--social, Berry's, Brownian, shear, and gravitational kernels. Coagulating systems consisting of a thousand monomers were observed in the late time of the evolution, for the moment when only fifty clusters were left in the system. In this way, the impact of particular initial condition was revealed in relation to other initial conditions. Several similarities between different kernels were observed, among others, strong correspondence between the constant and the Brownian kernel was confirmed. In case of the product kernel, unexpected behaviour related to the condensation nuclei initial conditions was revealed and explained. According to the results, it is doubtful that a phase transition can be triggered by appropriate setting of initial condition. 
\end{abstract}

\maketitle

\section{Introduction}\label{SectIntro}

Aggregation (also known as coalescence or coagulation) is a process of merging clusters irreversibly into larger clusters. Usually, we consider binary aggregation acts only, so no more than two clusters can join at a time. Such an assumption makes the aggregation process much more tractable and, in real, can be imagined as a system diluted enough to prevent three--body collisions. The reaction scheme is then
\begin{equation} \label{general_scheme}
   \left(i\right)+\left(j\right){{\stackrel{K\left(i,j\right)}{\longrightarrow}}}\left(i+j\right),
\end{equation}

\noindent where $\left(i\right)$ stands for a cluster of mass $i$ and $K\left(i,j\right)$ is the coagulation kernel representing the rate of the process (reaction rate). The system is closed, thus, eventually, all of the clusters join to form a single giant cluster.

Such an abstract definition of aggregation causes that a large number of everyday--life processes can be classified as it. Very first examples are blood coagulation, milk curdling or water vapor condensation but a lot of others can be recalled: formation of traffic jams, gravitational formation of planets (accretion), or even consolidation of companies known from corporate finance. Another important example is water treatment, in which dispersed colloidal particles agglomerate (flocculation). At this point, it is worth to emphasize that $i$ and $j$ in Eq.~(\ref{general_scheme}) not necessarily must mean mass but also size, capitalization or other additive quantity. 

Despite its ubiquity and its relatively simple definition, we are far from a complete theory of aggregation. Originally, it was studied by Smoluchowski \cite{Smoluchowski}. Later, several authors investigated aggregation for formation of polymers \cite{Stockmayer_1943}, aerosols \cite{aerosols}, or food and material processing \cite{1981_Schmidt, Wattis_2004}. In the last two decades, aggregation has been applied to a broad range of interdisciplinary topics, including percolation phenomena in random graphs and complex networks \cite{2005_JPhysALushnikov, 2009_ScienceAchlioptas, 2010_PRLCosta, 2010_PRECho, 2016_PRLCho, 2016_Conv}, pattern formation \cite{2002_PREDubovik,2007_EcolModelSaadi, 2014_PREMatsoukas, 2014_SciRepMatsoukas}, and population genetics \cite{2005book_Hein}. Recently, some physiological processes were modeled using aggregation formalism \cite{2020_PRE_Nelson, 2020_Miangolarra}.

There are three basic approaches to investigate aggregation: experimental, theoretical, and simulational. An excellent example of a straightforward experiment is milk curdling. However, it also shows the main problem with experimental aggregation: it is very difficult to track sizes of all or at least of most of the particles that take part in the process. The heuristic result that the substance coagulated and its properties changed is obvious but the most interesting information (from statistical physics point of view) -- cluster size distributions in subsequent moments of time -- remain unknown. Nevertheless, for some special types of systems it is possible to track cluster size distribution during the coagulation. Electrorheological fluid is the example of such a system. In this case, statistics of clusters of particular size were calculated using a microscope and a camera \cite{Mimouni}.

There exist several theoretical approaches to aggregation. The classic approach is based on the Smoluchowski aggregation equation. A condensed overview of that can be found in \cite{Krapivsky}. Some number of kernels were solved explicitly in Smoluchowski formalism (by solving a kernel we mean predicting cluster size distribution for any moment of the process), some others using approximations or some information on their behaviour were derived by grouping them into so--called universality classes of kernels of similar properties. Smoluchowski model is widely used, although, solution to the Smoluchowski equation may not necessarily be the same as solution to the coagulation process itself. Further comment on that issue can be found in \cite{2018_PREFronczak}. Nevertheless, classic approach is widely used and researched \cite{2020_PRE_Nelson, 2020_Miangolarra, 2021_Leyvraz, 2021_Velasquez}.

The second theoretical approach was proposed by Marcus \cite{Marcus} and was based on the master equation. Opposite to the Smoluchowski formalism, Marcus approach did not use mean values of cluster concentrations but temporary cluster size distributions. Due to this fact, solutions in this approach are exact solutions to the coagulation processes. Later, Lushnikov gave a significant contribution to that formalism \cite{2004_PRLLushnikov, 2005_PRELushnikov, 2011_JPhysALushnikov}. However, it requires involved mathematics and only a few basic kernels were solved there.  

Recently, several basic as well as non--trivial kernels were solved in the combinatorial approach \cite{grassberger1, grassberger2, 2019_ROMP_Lepek, 2021_PhysD_Lepek, 2021_ROMP_Lepek}. In this case, combinatorial expressions were derived to obtain space of available states for the coagulating system. 

Theoretical (mathematical) description of aggregating systems is of the highest interest as it allows to predict states (cluster size distributions) of the system at particular points in time and to derive some general conclusions on its behaviour which can be compared to other types of systems. However, numerical simulations of aggregating systems are not less useful. Numerical (computer) simulation can be regarded as a source of truth and as a reference point for theoretical considerations. It is due to the fact that we can store all of the clusters in the computer memory and update the system state step-by-step by choosing randomly (with appropriate $K$) clusters to be merged. Such a computer program is an exact realization of the aggregation definition. Of course, because of the randomness we must always collect enough number of independent program runs to obtain reliable statistics. For details of simulations, see Section \ref{SectSim}.

\begin{center}
$\diamondsuit$    
\end{center}

For some kernels, a giant (macroscopic, infinite) cluster appears during the evolution of the system. Obviously, physical properties of the system changes dramatically. Those kernels are called gelling, in opposition to non--gelling kernels where giant clusters do not arise. The most famous gelling kernel is the product kernel, $K \propto ij$, but there are also other forms as powers of kernels, $(ij)^2$, $(i+j)^2$, or condensation kernel, $(i+A)(j+A)$. Such a sol--gel transition is generally considered as of the second kind (continuous phase transition) \cite{2005_PRELushnikov}.

Typically, aggregating systems are researched starting from the monodisperse initial conditions -- all clusters at the beginning of the process are of size one. It is due to the fact that there were very few successful attempts to theoretical description of systems which coagulation starting from other initial conditions. There exists evidence, however, that coagulating system behaviour may depend on its initial condition \cite{2004_Menon}. An example of solution for the product kernel in Marcus--Lushnikov approach can be found in \cite{2019_PRE_Fronczak}. Another example is the solution for the constant kernel (in Smoluchowski approach) with algebraically decaying initial conditions from \cite{Krapivsky} (p. 143). Unfortunately, it has been shown before that Smoluchowski solutions are good approximations only for early stages of the system's evolution \cite{2019_ROMP_Lepek}. Until now, as far as I am concerned, there were no systematic studies in this topic, neither theoretical nor by simulation. Taking into account that virtually none of the kernels were solved theoretically with arbitrary initial conditions, a broad numerical study can be considered unavoidable.

Previous works gave insight on potentially interesting initial conditions \cite{2019_PRE_Fronczak}. These are: exponential initial conditions, $n_s^{t=0} \propto e^{-\lambda s}$, and power--law initial conditions, $n_s^{t=0} \propto s^{-\alpha}$. Here, $ n_s^{t} $ stands for the number of clusters of size $s$ at time $t$. The latter are of special importance as power--law distributions arising in the critical point are one of the signs of continuous phase transition \cite{Yeomans, MEJNewmanParetoLaw}. Additionally, I will test initial conditions consisting of a few large clusters and the rest of clusters of size one. Such a condition is inspired by cloud condensation nuclei which are small particles on which water vapor condenses (I will call this initial condition as condensation nuclei, CN). To sum up, what is the dependence on the initial condition? Is it possible to drive an aggregating system into a phase transition by setting up appropriate initial condition? In this work, I try to answer this questions by examining several types of kernel forms.

Readers interested in further literature on aggregation can consult \cite{paper2,paper3,paper7,paper8,paper9, Aldous}.

\section{Numerical simulation}\label{SectSim}

As mentioned in the Introduction, by a numerical simulation I mean a computer program which holds in its memory labelled clusters existing currently in the system, and their sizes. In the subsequent steps the system state is updated by choosing randomly two clusters to be merged. This random choosing is performed proportionally to the kernel $K(i,j)$, i.e., the probability of choosing clusters of sizes $i$ and $j$ is proportional to $K$. An algorithm representing overall simulation with any kernel is presented as Algorithm 1 and was adopted from \cite{eibeck2000}.

\label{sec:Annex1}
	\begin{algorithm}[H]  
		\caption{Simulation of the coagulation process with arbitrary kernel} 
		\begin{algorithmic}[1] 
			\Require $N$ - no. of monomers, $N_0$ - no. of initial clusters, $\mathbb{V} =\left\lbrace v_i\mid i=1,2,...,N_0\right\rbrace $ - a sequence of initial clusters' sizes, $maxtime$ - time of simulation
			\Ensure Cluster sizes at time $t$
			\Function {Coagulation}{$\mathbb{V}$, $maxtime$} 
			\State $t\gets t_{start}$
			\While{TRUE}
			\State $t\gets t+1$
			\If{$t>maxtime$}
			\State $break$
			\EndIf
			\State draw two distinct clusters $\left\lbrace v_i,v_j\right\rbrace$ in a way that probability of choosing the pair $\left\lbrace v_i,v_j\right\rbrace$ is proportional to $K(i,j)$  
			\State $\mathbb{V} \gets \mathbb{V} \cup \left\lbrace v_i+v_j \right\rbrace$ \Comment{add a new (merged) cluster}
			\State $\mathbb{V} \gets \mathbb{V} \setminus \left\lbrace {v_i,v_j} \right\rbrace$ \Comment{remove two old clusters}
			\EndWhile
			\State \Return{$\mathbb{V}$}  
			\EndFunction  
		\end{algorithmic}  
	\end{algorithm} 

\begin{figure} 
\includegraphics[scale=0.63]{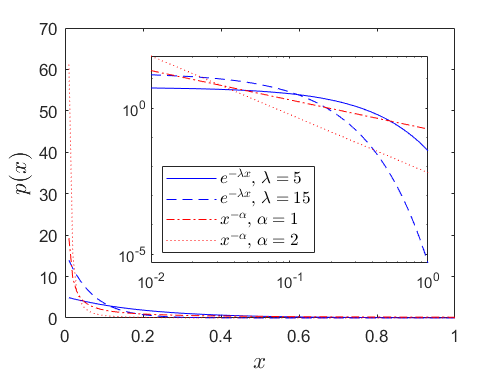}
\caption{Distributions used for the generation of initial cluster sizes. Two functions were used: exponential, $e^{-\lambda x}$, and power--law, $x^{-\alpha}$. Due to practical reasons, for $x>1$, we require a sharp cut--off, $p(x>1)=0$. Distributions are scaled due to the condition $\int_0^1 p(x)dx=1$ to be convenient to compare. Inset figure: the same data in log--log scale.}
\label{Figure_distributions}
\end{figure}

\begin{figure}
\includegraphics[scale=0.63]{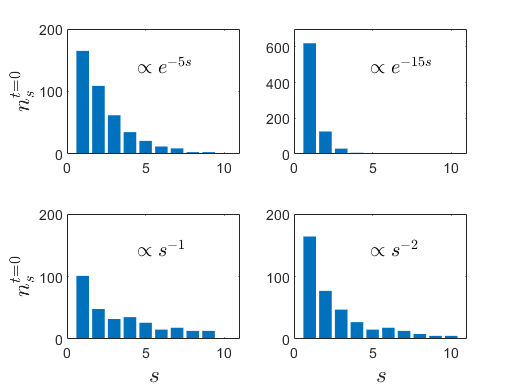}
\caption{Examples of initial cluster size configurations, $n_s^{t=0}$, generated with exponential (upper plots) and power--law (lower plots) distributions. The initial configuration was limited to generate cluster sizes from 1 to 10, thus, we have $n_{s}^{t=0}=0$ for $s>10$. The were also CN initial conditions used in this study. One realization of CN was a set of ten clusters of size 10 and the rest (nine hundred) of size 1. The second realization consisted of one single condensation nucleus containing 0.1 of the system's mass and nine hundred clusters of size 1. For all cases, the total number of monomers in the system was $N=1000$, so the sum of initial cluster sizes was 1000 as well.}
\label{Figure_initial_examples}
\end{figure}

Vector $\mathbb{V}$ in Algorithm 1 is the vector containing initial sizes of clusters. In case of the monodisperse initial conditions, $\mathbb{V}$  contains ones only. Other initial conditions are generated by modifying $\mathbb{V}$, namely, by generating randomly this sequence due to chosen distribution. In case of the exponential initial conditions, the histogram of elements $v_i$ (cluster size) shall behave as exponent. By analogy, for power--law initial conditions, the histogram of $v_i$'s shall behave as power--law function. The order of elements $v_i$ does not matter because later random choosing (line no. 8) depends only on the size of the cluster, not on the order. The code for that has been added to the previously used programistic library \cite{cpp_libraries}. Figure \ref{Figure_distributions} presents distributions used for initial data generation. In Figure \ref{Figure_initial_examples}, some realizations of these initial conditions are shown. Beside exponential and power--law conditions, there were also two types of CN initial conditions used: (i) consisting of ten nuclei of size 10 and nine hundred of size 1, and (ii) one single nucleus of size 100 (which is $0.1N$) and nine hundred nuclei of size 1.

In line no. 2 of the algorithm, we set time $t$ as $t_{start}$ which represents starting time. A thorough comment must be given on that. In this kind of simulation of the irreversibly aggregating system, we assume that at each time step $t$ a single binary merging act occurs. It means that the number of clusters, $k$, decreases linearly with time steps, $k=N-t$. Therefore, current $t$ represents actual stage of the system's evolution. For instance, for $t=0$, we have a situation in which there are $k=N$ separate clusters and none of the coagulation acts occurred (and this corresponds to monodisperse initial conditions). For $t=57$, we know that the system is 57 coagulation acts away from the monodisperse initial condition, and that there are $N-t$ steps to be performed until all the particles merge into one single cluster (which corresponds, of course, to time $t=N-1$).

Now, please notice that a particular initial condition corresponds to a particular value of $t$. For example, a single condensation nucleus initial condition (a single particle of size 100) corresponds to $t=99$ because 99 coagulation acts were needed to create such a state starting from monodisperse initial condition. Thus, I will call this starting value of $t$ as $t_{start}$ throughout the work. In case of random choosing (exponential and power--law initial conditions), $t_{start}$ must be calculated, e.g., due to the expression
\begin{equation} \label{t_start}
   t_{start} = \sum_{i=1}^k{( s_i - 1 )},
\end{equation}

\noindent where we sum masses of all of the clusters, $s_i$, decreased by one.

The sense of using $t_{start}$ is to make sure that all of the simulations finish at the same time step $t$ (e.g., $t=950$, corresponding to the state with $k=50$ clusters in the system of $N=1000$ monomers). It will allow us to directly compare the influence of particular initial conditions to the later stage of the system's evolution.

Line no. 8 in Algorithm 1 also needs a special explanation. The method to perform choosing two clusters proportionally to the kernel $K(i,j)$ has been thoroughly described in Appendix in \cite{2021_PhysD_Lepek}. Here, for clarity, I will recall the most crucial information.

\begin{table*}[ht!]
\begin{center}
\setlength\extrarowheight{2.5pt}
\setlength{\tabcolsep}{20pt}
\begin{tabular}{  l  l  }
\hline
\hline
$K(i,j)$ & Comment \\[2.5pt]
\hline
\hline
$\mathrm{const}$ & Constant \\[2.5pt]
\hline
$i+j$ & Additive  \\[2.5pt]
\hline
$ij$ & Product \\[2.5pt]
\hline
$ i^{-1}+j^{-1} $ & Electrorheological  \\[2.5pt]
\hline
$(i+j)^2(ij)^{-1}$ & Anti--social \\[2.5pt]
\hline
$(i-j)^2(i+j)^{-1}$ & Approx. Berry's kernel \\[2.5pt]
\hline
$(i^{1/3}+j^{1/3})(i^{-1/3}+j^{-1/3})$ & Brownian motion (continuum regime)  \\[2.5pt]
\hline
$(i^{1/3}+j^{1/3})^3$ & Shear (linear velocity profile) \\[2.5pt]
 \hline
$(i^{1/3}+j^{1/3})^2|i^{1/3}-j^{1/3}|$ & Gravitational settling  \\[2.5pt]
\hline
\end{tabular}
\end{center}
\caption{\small Kernels that have been used for this study. Further description in the text.}
\label{table}
\end{table*}

Let suppose that kernel expression we consider gives real numbers, e.g., $K(i,j)=\left( 1/i+1/j \right)$. This is the most general case that can occur.  We must construct a vector to store probabilities of all possible coagulation acts. Let me suppose that we have 3 particles in our system, A of size 1, B of size 2, and C of size 5. As we can calculate using the expression for $K$, the probability of merging together particles A and B is 1.5, the probability of merging A and C is 1.2, and the probability of merging B and C is 0.7. At this point, we construct a vector which contains cumulative sums of the calculated probabilities. Namely, in the first cell we put the probability of the merging event for pair A+B, in the second cell we put the sum of probabilities of A+B and A+C, and in the last cell we put the sum of probabilities of all pairs. Thus, our vector consists of three cells: [1.5, 1.5+1.2, 1.5+1.2+0.7], which, finally, is [1.5, 2.7, 3.4]. Now, we must randomly choose (i.e., using uniform distribution) a real number ranging from 0.0 to 3.4, i.e., from zero to the maximal (last) value in the vector. This random number will indicate one particular cell in the vector. For example, let us say we have randomly chosen the number of 1.9. We must check whether it is lower or equal to the value in the first cell (which is 1.5). Our random number is higher, thus, we go to the next cell. This time, the number of 1.9 is lower than the probability in the second cell of the vector (which is 2.7), therefore, it is the cell of interest. We considered that second cell as representing the pair of A+C, therefore, we shall take the pair A+C as the coagulation act for this particular step of our simulation.

Throughout the work, a system of $N=1000$ monomers (i.e., the smallest indivisible particles) is considered.

\section{Methods}\label{SectMeth}

A strong relation exists between the product kernel and percolation phenomenon in Erdos--Renyi networks \cite{Krapivsky}. In ER network, a search for the phase transition can be performed by tracking the ratio of number of nodes that contribute to the giant (percolating) cluster to the number of all nodes in the network, $r = n_{max} / N$, which is usually regarded as the order parameter of this system. However, an exact location of the critical point shall be determined by tracking another quantity, i.e., an average size of the second--biggest cluster or, alternatively, average size of all clusters without the giant one,  $\langle s_{-max} \rangle$. A description of methods for numerical research on phase transitions can be found in the book by Landau and Binder \cite{Binder_book} or in more comprehensive book by Newton and Barkema \cite{Newman_Monte_Carlo}. 

Average size of the second--biggest cluster in gelling systems behaves similar as magnetic susceptibility in magnetic systems \cite{Yeomans}. Before the critical point average size of the second--biggest cluster increases as same as average size of clusters in general. At the critical point, this quantity diverges (in the thermodynamic limit, i.e., for an infinite system) and, as evolution goes on, it decreases again because the giant cluster takes over the mass from the second--biggest cluster.

In this study, we will take advantage of the equivalence of gelation and ER network percolation \cite{2019_Souza}. Therefore, the ratio $r = n_{max} / N$, and the average size of the clusters other than giant one,  $\langle s_{-max} \rangle$, will be used to observe behaviour of the aggregating systems under various kernels and different initial conditions. Additionally, we will observe the average number of clusters of given size for the latest stage of the evolution when there are only $k=50$ clusters left in the system, namely $\langle n_s^{k=50} \rangle$. Any giant cluster will be recognizable on the plot.

\begin{figure*}
\includegraphics[scale=0.8]{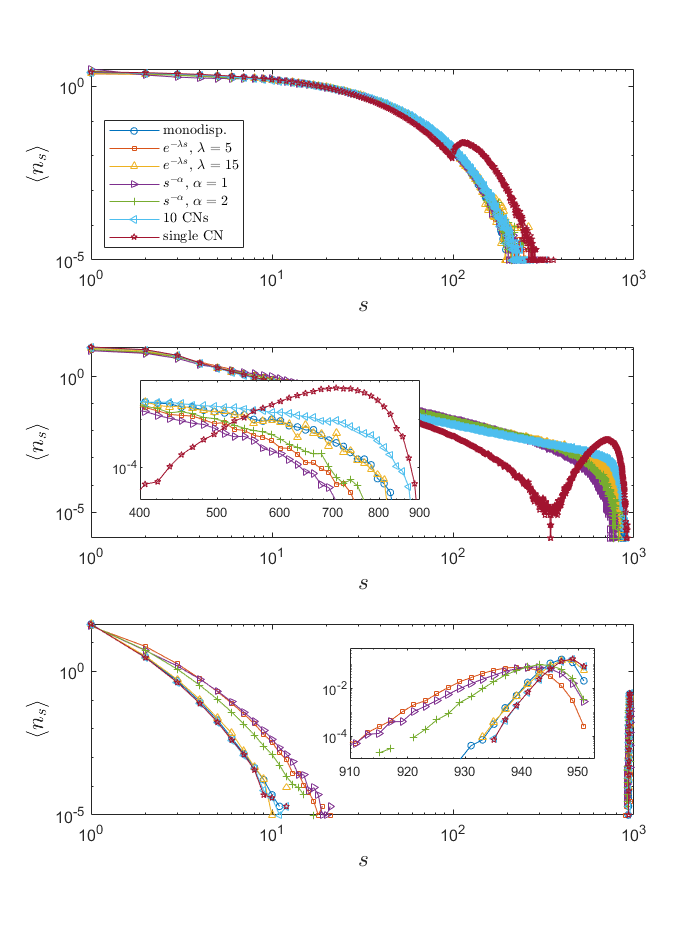}
\caption{Average number of clusters, $\langle n_s \rangle$, of given size $s$ for constant, $K=\mathrm{const}$ (upper plot), additive, $K=i+j$ (middle plot), and product, $K=ij$ (lower plot), kernels. Each data series represents particular initial conditions. These were: monodisperse (circles, blue), exponential with $\lambda=5$ (squares, red), exponential with $\lambda=15$ ($\triangle$, yellow), power--law with $\alpha=1$ ($\triangleright$, purple), power--law with $\alpha=2$ (crosses, green), ten condensation nuclei (CN) of size 10 ($\triangleleft$, light blue), and single condensation nucleus of size $0.1N=100$ (stars, brown). Figures represent a very late point in the system's evolution when only $k=50$ clusters are left in the system ($t=0.95N$). There were $10^5$ independent simulation runs performed for each data series. Legend is valid for all the plots. Further description in the text.}
\label{Figure_3}
\end{figure*}

\section{Kernels}\label{SectKern}

In this work, I examined the influence of initial conditions to several types of kernels. The kernels of interest are listed in Table \ref{table}. The basic kernels (constant, additive, and product) are used to show reference behaviour and the reference impact to the system caused by the initial conditions other than monodisperse ones. Electrorheological (ER) and anti--social kernels were previously studied in the combinatorial approach. The latter has an interesting feature of giving low probabilities for merging particles of similar sizes while giving high probabilities for merging particles of significantly different sizes \cite{Phd_thesis}. In the ER kernel, events with large clusters occur with very low probability. ER kernel is also interesting as it is one of few kernels that were thoroughly examined experimentally. Further kernels were adopted from the comprehensive study by Aldous \cite{Aldous}. They are: analytical approximation of Berry's kernel, Brownian motion kernel, shear kernel, and gravitational settling kernel. The Brownian kernel is particular example of the general kernel form defined by Fournier and Lauren\c{c}ot, $(i^\alpha+j^\alpha)(i^{-\beta}+j^{-\beta})$ \cite{Fournier_2005}. Gravitational settling and shear kernels are related to industrial processes.


\begin{figure*}
\includegraphics[width=\textwidth]{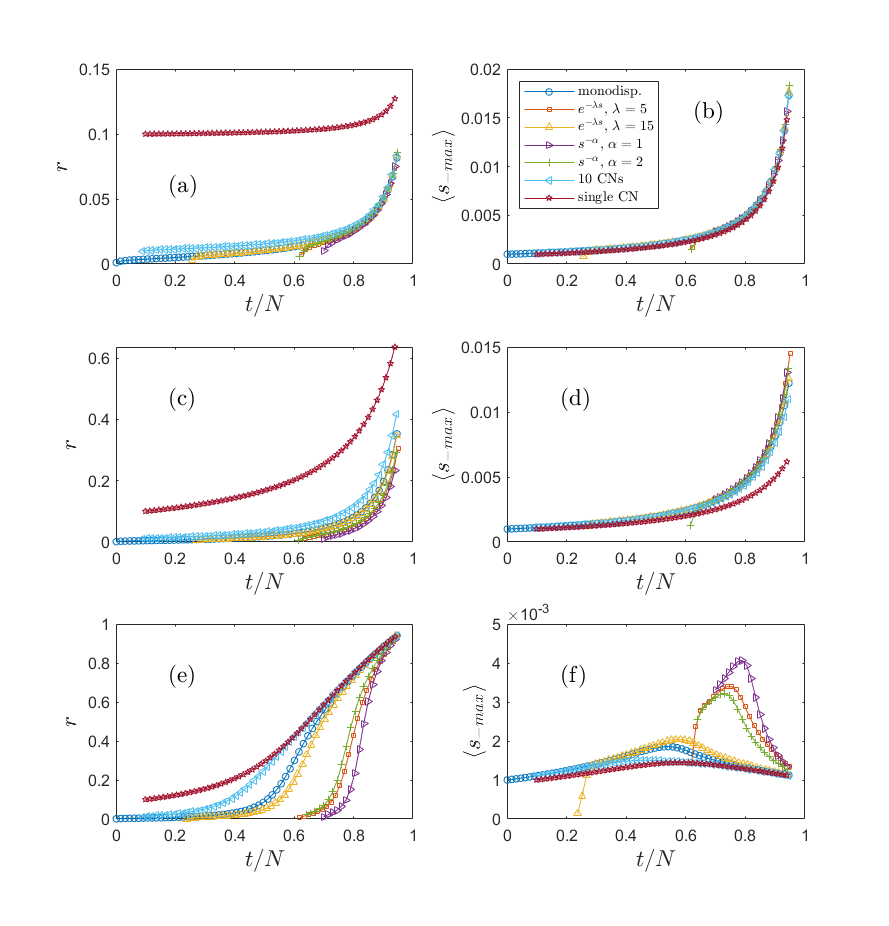}
\caption{The number of monomers included in the giant cluster, $r = n_{max} / N$ (left column), and average size of all clusters except for the giant one, $\langle s_{-max} \rangle$, (right column) for (a, b) constant, $K=\mathrm{const}$, (c, d) additive, $K=i+j$, and (e, f) product, $K=ij$, kernels. Each data series represents particular initial conditions. These were: monodisperse (circles, blue), exponential with $\lambda=5$ (squares, red), exponential with $\lambda=15$ ($\triangle$, yellow), power--law with $\alpha=1$ ($\triangleright$, purple), power--law with $\alpha=2$ (crosses, green), ten condensation nuclei (CN) of size 10 ($\triangleleft$, light blue), and single condensation nucleus of size $0.1N=100$ (stars, brown). Figures represent a very late point in the system's evolution when only $k=50$ clusters are left in the system ($t=0.95N$). There were $10^5$ independent simulation runs performed for each data series. Legend in (b) is valid for all the plots. Further description in the text.}
\label{Figure_4}
\end{figure*}

\begin{figure}
\includegraphics[scale=0.58]{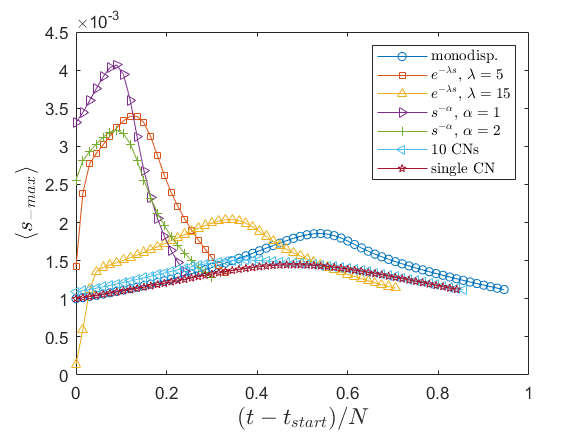}
\caption{Average size of finite clusters (all clusters except for the giant one), $\langle s_{-max} \rangle$, for the product kernel, $K=ij$. The series have been shifted by the factor of $t_{start}$. In this way, one can observe how fast after the start of the simulation a phase transition into gel occurred. The point of the phase transition is indicated by the maximum value of $\langle s_{-max} \rangle$, however, the respective values may vary for different system sizes $N$ (finite--size scaling). Here, the system size was $N=1000$ and each series was averaged over $10^5$ independent runs.}
\label{Figure_5}
\end{figure}

\section{Results}\label{SectRes}

For each kernel, I plotted the tracking quantities, $r$ and $\langle s_{-max} \rangle$, for monodisperse, exponential, power--law, and CN initial conditions. Monodisperse initial conditions played a role of the reference data. Additionally, for exponential and power--law initial conditions, several values of parameters $\lambda$ (5, 15) and $\alpha$ (1, 2) were tested. In all figures, data points from numerical simulation are depicted with markers and continuous lines are only guidelines for eyes.

Figure~\ref{Figure_3} presents average number of clusters, $\langle n_s \rangle$, of given size $s$ for constant (upper plot), additive (middle plot), and product (lower plot), kernels. We observe a very late point in the system's evolution when only $k=50$ clusters are left in the system (i.e., $t=0.95N$). Each data series represents particular initial condition. In case of the constant kernel (upper plot), most of the data series (except for single CN initial) overlap, therefore, initial condition does not impact the process at all. It stays in perfect compliance with predictions as in this case the probability of merging two clusters does not depend on its size, thus, it neither depends on initial configuration of cluster sizes. For single CN initial condition, the system started from the state where there were nine hundred clusters of size 1 and one cluster of size 100. This single big cluster grew as same a other clusters and this is the reason why we observe a slight peak of $\langle n_s \rangle$ for $s>10^2$. Although a distinct peak in cluster size distribution on the right--hand side of the minimum is often related to gel phase (cf. Fig. 1 in \cite{2019_PRE_Fronczak}), it is not the case here what we may instantaneously conclude looking at Figure \ref{Figure_4}b (no unimodal maximum).

The middle plot in Figure~\ref{Figure_3} presents results for the additive kernel. In this case, the impact of single CN initial condition to the system is similar as for the constant kernel but even more intense. Here, again, the distinct peak in $\langle n_s \rangle$ is not a sign of phase transition as the average size of finite size clusters in Figure~\ref{Figure_4}d is a monotonically increasing curve. Opposite to the previous example of the constant kernel, various initial conditions showed impact on the final system state resulting in different maximum sizes of cluster that \textit{could have grown} in the system ($s_{max}$ from 700 to 900).
The biggest clusters were obtained for single CN initial condition. The next were monodisperse and exponential ($\lambda=15$) initial conditions. The similarity in results between these two cases shall not be the surprise as the initial cluster size configurations were similar (cf. Figure 2). Interestingly, power--law initial conditions resulted in relatively small maximal cluster sizes in the system at $t=0.95N$. How could that happen? Bear in mind that different initial conditions are related to different values of $t_{start}$: for monodisperse conditions, $t_{start}=0$ but, for power--law initial conditions, $t_{start}$ is a high number depending on a particular realization. For instance, in Figure~\ref{Figure_4}d, we can see that plot for power--law conditions with $\alpha=2$ starts from $t$ slightly higher than 600. Thus, this unexpected observation that power--law initial conditions produce smaller maximal cluster sizes at $t=950$ may be due to the fact that when system evolves from monodisperse initial conditions then at time close to $t_{start}$ for power--law conditions there are more large clusters than in the case of starting from power--law conditions (where initial configuration is limited to sizes from 1 to 10, cf. Figure 2).

The above observation is valid also for the product kernel (Figure 3, lower plot). Here, we can see that power--law initial conditions and exponential initial conditions with $\lambda=5$ resulted in smaller possible clusters at time of observation, $t=950$, than the rest of initial conditions. Also, again, CN initial conditions generated the largest possible clusters.

The product kernel is a gelling one what can be clearly seen from Figure 4f. Figure 4 was mentioned above but I will now describe it more thoroughly. It presents the number of monomers included in the giant cluster, $r = n_{max} / N$ (left column), and average size of all clusters except for the giant one, $\langle s_{-max} \rangle$, (finite--size clusters, right column). The data in Figure 4 start at the beginning of numerical simulation ($t_{start}$) and ends for $t=950$ when the number of clusters is $k=50$. Phase transition occurring during the product kernel aggregation is visible in form of unimodal peak in $\langle s_{-max} \rangle$. The data series in Figure 4 start from different values of $t_start$ due to the issue described before. 

For the gelling product kernel, it would be interesting to ask a question how fast the phase transition occurs after the start of the simulation (not for what values of $t$) for different initial conditions? To answer that, let us look into Figure 5 where I have shifted the data to plot $\langle s_{-max} \rangle$ versus $(t-t_{start})/N$. The point of the phase transition is indicated by the maximum value of $\langle s_{-max} \rangle$, however, the respective values may vary for different system sizes $N$ (finite--size scaling; here, I used $N=1000$). The shortest time from the start of the simulation to the time of the phase transition was observed for power--law initial conditions with $\alpha=1$ and it was $t-t_{start}=t_c = 87$. For other conditions, the times $t_c$ were: power--law ($\alpha=2$): $t_c=88$, exponential ($\lambda=5$): $t_c=127$, exponential ($\lambda=15$): $t_c=342$, 10 CNs: $t_c=376$, single CN: $t_c=480$, and monodisperse: $t_c=538$.

Figure 6 presents average number of clusters of given size, $\langle n_s \rangle$, for electrorheological (ER for short, upper plot), anti--social (middle), and approximation of Berry's kernel (lower). Each data series represents particular initial conditions: monodisperse (circles, blue), exponential with $\lambda=5$ (squares, red), exponential with $\lambda=15$ ($\triangle$, yellow), power--law with $\alpha=1$ ($\triangleright$, purple), power--law with $\alpha=2$ (crosses, green), 10--CNs condition  ($\triangleleft$, light blue), and single CN of size $0.1N=100$ (stars, brown). In case of the Berry's kernel, there are no data for monodisperse initial condition as evolution could not start ($K=0$ for all of the clusters). In Figure 6, as same as in Figure 3, plots represent late point in the system's evolution when only $k=50$ clusters are left in the system ($t=0.95N$). There were $10^4$ independent simulation runs performed for each data series.

The results show that in cases of the ER and anti--social kernels initial conditions do not impact evolution in a significant way. The only outlier is the case of single CN. For both ER and anti--social kernels, there is a fraction of large-size clusters on the right--hand side of the distribution. However, similarly as for the additive kernel, there is no sign of phase transition into gel (Figure 7b and 7d), thus this fraction is only the result of the original giant nucleus growing. For ER kernel, it is especially clearly visible as ER kernel generates very low probabilities of merging $K$ for large clusters, thus the original peak for $s=100$ is still visible.

Lower plot in Figure 6 presents data for the approximation of Berry's kernel. Here, power--law cluster size distributions were produced for exponential and power--law initial conditions. For ten CNs and single CN, initial monomers could only merge with the CNs. Ten CNs initial condition resulted in large number of big clusters (peak for $s=100$). Single CN initial condition resulted only in two data points, due to the fact that for Berry's kernel, $K=(i-j)^2(i+j)^{-1}$, encounters between monomers are not possible ($K=0$), therefore monomers can only join the giant nucleus.

Figure 7 contains respective plots of the number of monomers included in the giant cluster, $r = n_{max} / N$ (left column), and average size of all clusters except for the giant one, $\langle s_{-max} \rangle$, for ER, anti--social, and Berry's kernels. In Figure 7e and 7f, a specific result for single CN initial condition is again due to the fact encounters other than with the largest cluster were forbidden.

Figure 8 is analogous to previous Figures 3 and 6 but contains data for brownian, shear, and gravitational kernel. Taking a look into upper plot of Figure 8 we can see that the brownian kernel behaves similar as the constant kernel, both qualitatively and quantitatively. In \cite{Krapivsky}, we read that the constant kernel is an approximation of the brownian kernel as both are constant--independent in terms of cluster size multiplication, $K(ai,aj)=K(i,j)$. Indeed, the approximation works even better than one might expect because the output is virtually the same for the late stage of the evolution under respective initial conditions.

The additive kernel and the shear kernel stay in analogous relation. Results of the shear kernel evolution (Figure 8, middle) are both qualitatively and quantitatively similar to those obtained for the additive kernel, including generation of power--law cluster size distributions and minimum for $s=200$ and single CN initial condition (although this valley is relatively shallow).

Lower plot in Figure 8 presents the gravitational kernel. In this case, we can see clear analogy to the Berry's kernel. The only viable difference are numbers of clusters of the largest size for 10 CN initial condition: here, they are larger than in the case of the Berry's kernel (namely, from $s=500$ to $s=950$).

Figure 9 contains respective plots of the number of monomers included in the giant cluster, $r = n_{max} / N$ (left column), and average size of all clusters except for the giant one, $\langle s_{-max} \rangle$, for brownian, shear, and gravitational kernels.

\begin{figure*}
\includegraphics[scale=0.8]{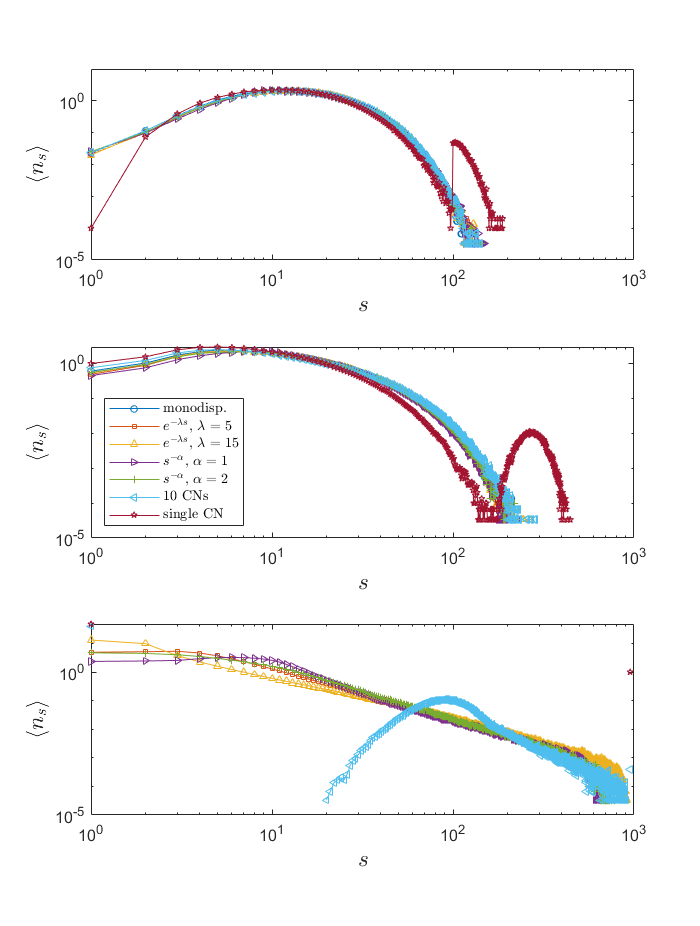}
\caption{Average number of clusters, $\langle n_s \rangle$, of given size $s$ for electrorheological, $K=i^{-1}+j^{-1}$ (upper plot), anti--social, $K=(i+j)^2(ij)^{-1}$ (middle), and Berry's kernel, $K=(i-j)^2(i+j)^{-1}$ (lower). Each data series represents particular initial conditions. These were: monodisperse (circles, blue), exponential with $\lambda=5$ (squares, red), exponential with $\lambda=15$ ($\triangle$, yellow), power--law with $\alpha=1$ ($\triangleright$, purple), power--law with $\alpha=2$ (crosses, green), ten condensation nuclei (CN) of size 10 ($\triangleleft$, light blue), and single condensation nucleus of size $0.1N=100$ (stars, brown). In case of the Berry's kernel, there are no data for monodisperse initial condition. Figures represent a very late point in the system's evolution when only $k=50$ clusters are left in the system ($t=0.95N$). There were $10^4$ independent simulation runs performed for each data series. Legend is valid for all the plots. Further description in the text.}
\label{Figure_6}
\end{figure*}

\begin{figure*}
\includegraphics[width=\textwidth]{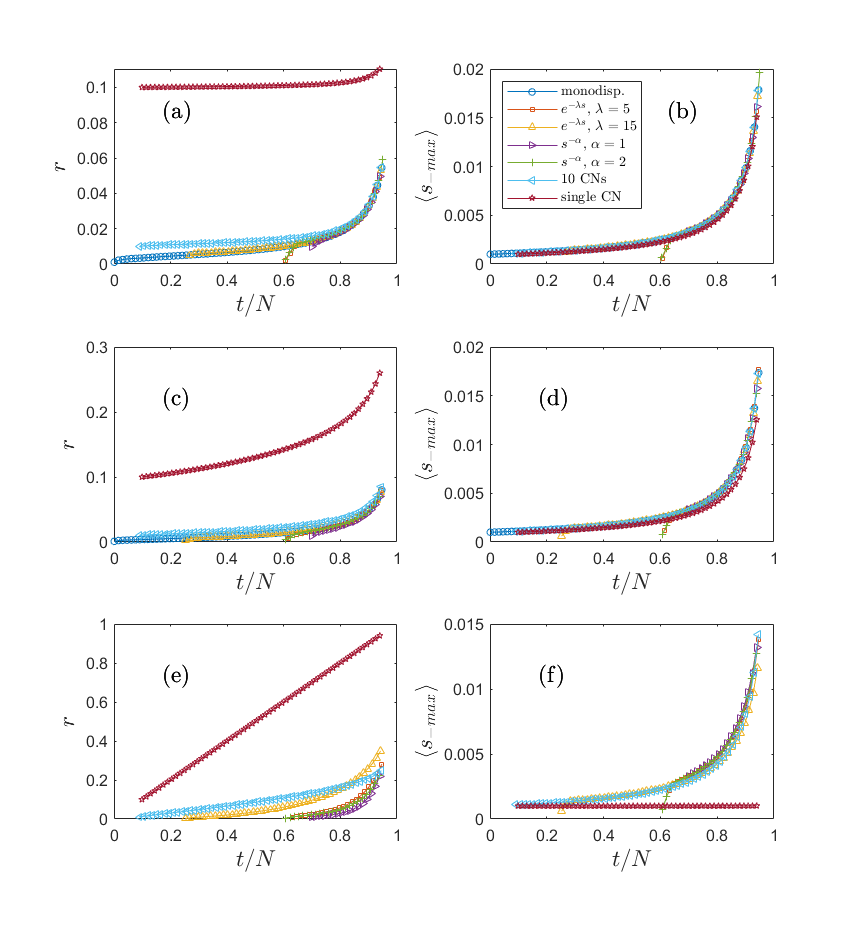}
\caption{The number of monomers included in the giant cluster, $r = n_{max} / N$ (left column), and average size of all clusters except for the giant one, $\langle s_{-max} \rangle$, (right column) for (a, b) electrorheological, $K=i^{-1}+j^{-1}$, (c, d) anti--social, $K=(i+j)^2(ij)^{-1}$, and (e, f) Berry's kernel, $K=(i-j)^2(i+j)^{-1}$. Each data series represents particular initial conditions. These were: monodisperse (circles, blue), exponential with $\lambda=5$ (squares, red), exponential with $\lambda=15$ ($\triangle$, yellow), power--law with $\alpha=1$ ($\triangleright$, purple), power--law with $\alpha=2$ (crosses, green), ten condensation nuclei (CN) of size 10 ($\triangleleft$, light blue), and single condensation nucleus of size $0.1N=100$ (stars, brown). In case of the Berry's kernel, there are no data for monodisperse initial condition. Figures represent a very late point in the system's evolution when only $k=50$ clusters are left in the system ($t=0.95N$). There were $10^4$ independent simulation runs performed for each data series. Legend in (b) is valid for all the plots. Further description in the text.}
\label{Figure_7}
\end{figure*}

\section{Conclusions}\label{SectConc}

For this work, several initial conditions were applied to coagulation processes with various kernels. Coagulating systems were observed in the late time of the evolution, namely, for the moment when only $k=50$ clusters were left in the system. In this way, the impact of particular initial condition was shown in relation to other initial conditions.

One of the most important results in the above work is probably Fig. 3b presenting additive kernel coagulation. Single CN initial condition (one cluster of size $s=100$ and the rest of clusters of size $s=1$) clearly stands out from the others. Qualitatively, the valley in the cluster size distribution tends to be similar to those observed for gelling kernels. However, a phase transition was not a case here, as the initial single CN cluster simply grew together with the other clusters in the system. This may be a noticeable observation as the inflection point in $\langle n_s \rangle$ was usually associated with crossing gelling point.

So may be that requiring the single CN initial condition ``sets'' the system in the state after the phase transition? It is doubtful as the average size of finite clusters increases for the whole time of the evolution, therefore, there is no sign of being after the transition point.

Another meaningful observation was made in Fig. 5. The time between the start of the simulation and the phase transition point was tested against several initial conditions. The result was that the most immediate transition occurred for power--law initial conditions. It stays in compliance with the fact that power--law distribution are usually considered as a sign of criticality. What was struggling there, was the fact that for coagulation starting from 10 CNs and single CN initial conditions, it took much longer time to achieve phase transition point. That time for single CN was close to the respective time for monodisperse initial conditions which may seem unexpected or even erroneous but to my best knowledge -- it is true. Power--law cluster size distribution is the most effective state to quickly build up a giant (percolation) cluster. When starting from monodisperse initial conditions, it takes a large number of steps to build up such a cluster. In case of the CN initial conditions, large clusters immediately intercept other (small) clusters so it is very hard for the second--largest cluster to build up and then decrease in size which would indicate the transition.

According to \cite{Krapivsky}, the constant kernel may be considered as approximation of the brownian kernel (coagulation equations for the Brownian kernel remain unsolved even for the simplest monodisperse initial condition). They both are invariant under the transformation $(i,j) \rightarrow (ai,aj)$, thus, authors state that ``the constant kernel is a reasonable (albeit uncontrolled) approximation of the physically important Brownian kernel''. Although in \cite{Krapivsky} this similarity was described mainly in reference to monodisperse initial conditions, by the above results we have shown that the approximation works better than well and correspondence between these two kernels is quantitatively very good for all of the researched initial conditions. 

Analogously, there are significant similarities in impact of different initial conditions to the additive kernel and shear kernel. Berry's kernel and gravitational kernel may be considered as another pair. These results may suggest that it is possible to efficiently approximate one kernel form by another, perhaps less complex and more susceptible to analytical approach, with respect to influence of initial conditions.

It seems that simulational studies on the coagulation phenomena remain to some extent an unexplored field, taking into account that any kernel form and any initial condition may be the subject of simulation. I believe that such computer--aided studies will help in research on real systems which are usually hard to be observed microscopically in experiment, namely, observation of cluster size distributions in real--time is often an issue (just to mention milk curdling).


\begin{figure*}
\includegraphics[scale=0.8]{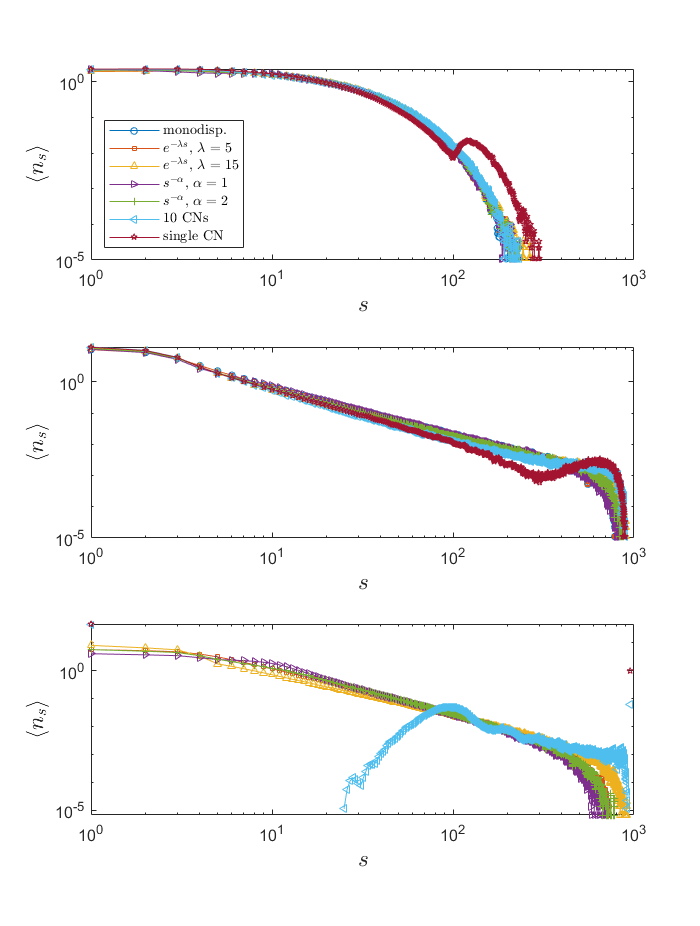}
\caption{Average number of clusters, $\langle n_s \rangle$, of given size $s$ for brownian, $K=(i^{1/3}+j^{1/3})(i^{-1/3}+j^{-1/3})$ (upper plot), shear, $K=(i^{1/3}+j^{1/3})^3$ (middle), and gravitational kernel, $K=(i^{1/3}+j^{1/3})^2|i^{1/3}-j^{1/3}|$ (lower). Each data series represents particular initial conditions. These were: monodisperse (circles, blue), exponential with $\lambda=5$ (squares, red), exponential with $\lambda=15$ ($\triangle$, yellow), power--law with $\alpha=1$ ($\triangleright$, purple), power--law with $\alpha=2$ (crosses, green), ten condensation nuclei (CN) of size 10 ($\triangleleft$, light blue), and single condensation nucleus of size $0.1N=100$ (stars, brown). In case of the gravitational kernel, there are no data for monodisperse initial condition. Figures represent a very late point in the system's evolution when only $k=50$ clusters are left in the system ($t=0.95N$). There were $10^4$ independent simulation runs performed for each data series. Legend is valid for all the plots. Further description in the text.}
\label{Figure_8}
\end{figure*}

\begin{figure*}
\includegraphics[width=\textwidth]{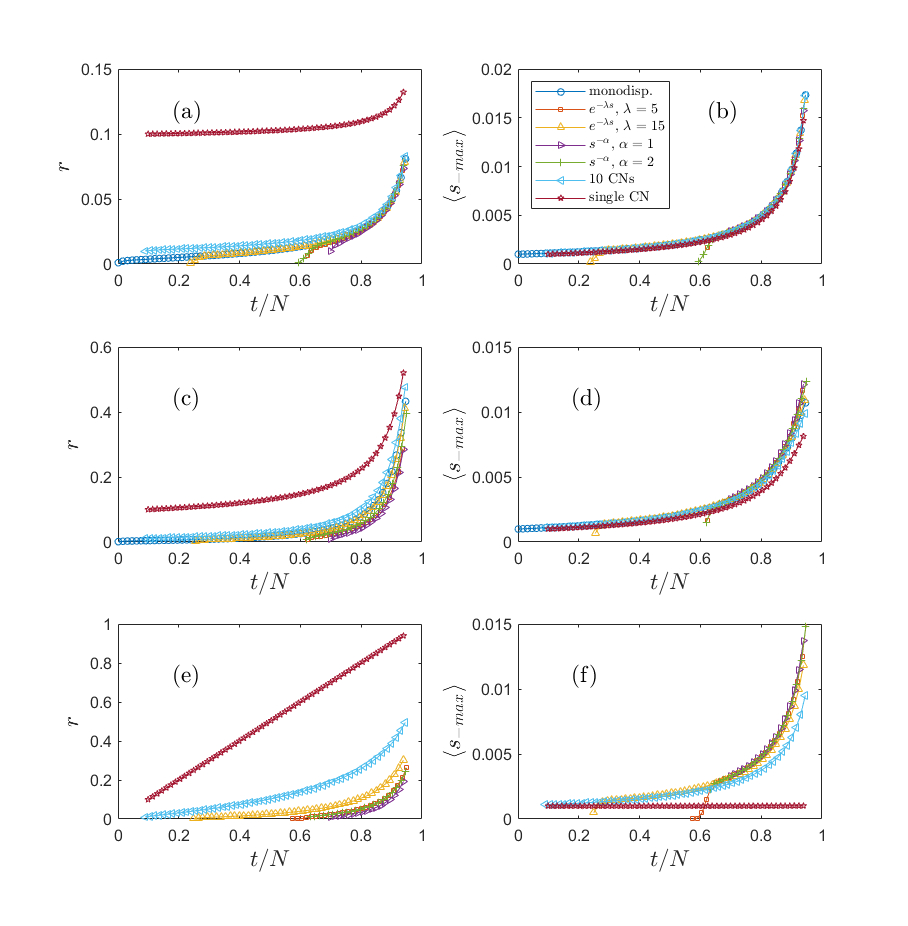}
\caption{The number of monomers included in the giant cluster, $r = n_{max} / N$ (left column), and average size of all clusters except for the giant one, $\langle s_{-max} \rangle$, (right column) for (a, b) brownian, $K=(i^{1/3}+j^{1/3})(i^{-1/3}+j^{-1/3})$, (c, d) shear, $K=(i^{1/3}+j^{1/3})^3$, and (e, f) gravitational kernel, $K=((i^{1/3}+j^{1/3})^2|i^{1/3}-j^{1/3}|$. Each data series represents particular initial conditions. These were: monodisperse (circles, blue), exponential with $\lambda=5$ (squares, red), exponential with $\lambda=15$ ($\triangle$, yellow), power--law with $\alpha=1$ ($\triangleright$, purple), power--law with $\alpha=2$ (crosses, green), ten condensation nuclei (CN) of size 10 ($\triangleleft$, light blue), and single condensation nucleus of size $0.1N=100$ (stars, brown). In case of the gravitational kernel, there are no data for monodisperse initial condition. Figures represent a very late point in the system's evolution when only $k=50$ clusters are left in the system ($t=0.95N$). There were $10^4$ independent simulation runs performed for each data series. Legend in (b) is valid for all the plots. Further description in the text.}
\label{Figure_9}
\end{figure*}

\acknowledgments
This work has been supported by the National Science Centre of Poland (Narodowe Centrum Nauki) under grant no.~2015/18/E/ST2/00560.

\end{document}